\documentstyle[aps,prc,preprint,tighten,epsf,floats]{revtex}
\let\lsim\lesssim

\let\citen\CITE

\begin{document}
\preprint{KUNS-1528}

\title{
Semiclassical Origin of Superdeformed Shell Structure in the
Spheroidal Cavity Model\thanks{%
Submitted to Progress of Theoretical Physics.}
}

\author{Ken-ichiro Arita}
\address{
Department of Physics, Nagoya Institute of Technology,
Nagoya 466-8555, Japan}

\author{Ayumu Sugita and Kenichi Matsuyanagi}
\address{
Department of Physics, Graduate School of Science,
Kyoto University, \\ Kyoto 606-8502, Japan
}

\date{September 2, 1997}
\maketitle

\begin{abstract}
Classical periodic orbits responsible for emergence of the
superdeformed shell structures for single-particle motions in
spheroidal cavities are identified and their relative contributions to
the shell structures are evaluated.  Both prolate and oblate
superdeformations (axis ratio about 2:1) as well as prolate
hyperdeformation (axis ratio about 3:1) are investigated.  Fourier
transforms of quantum spectra clearly show that three-dimensional
periodic orbits born out of bifurcations of planar orbits in the
equatorial plane become predominant at large prolate deformations,
while butterfly-shaped planar orbits bifurcated from linear orbits
along the minor axis are important at large oblate deformations.
\end{abstract}

\section{Introduction}
\label{sec:intro}

In the last decade, superdeformed spectroscopy, i.e., study on nuclear
structure with large prolate deformations (axis ratio about 2:1), has
developed enormously, and further significant progress is
expected\cite{twin,khoo,gamma}.

It is well known that the superdeformation is a result of deformed
shell effect and in fact realistic calculations of both
Strutinsky-Nilsson type and Hartree-Fock type work well for discussing
shell structures observed in experiments at such large
deformations\cite{aberg}.  The purpose of this paper, however, is not
to make some realistic calculations in relation to recent experimental
findings.  Rather, we address here to the fundamental question ``why
nucleus is superdeformed'' and investigate semiclassical origin of
emergence of the superdeformed shell structure in a simple model,
i.e., spheroidal cavity model.

According to the periodic-orbit theory\cite{gutw,balian,BT,BB} based
on the semiclassical approximation to the path integral, oscillating
parts of single-particle level densities coarse-grained with respect
to a certain energy resolution, i.e., shell structure we are
interested in, are determined by classical periodic orbits with short
periods.  As is well known, nucleus favors such shapes at which
prominent shell structures are formed and its Fermi surface lies in a
valley of the oscillating level density, increasing its binding energy
in this way.

With the semiclassical approach, Strutinsky et al.\cite{stru} studied
the shell structure associated with the spheroidal cavity model and
found that planar orbits in the meridian plane are responsible for the
shell structure at normal prolate deformations.  In addition, they
pointed out that some three-dimensional (3D) periodic orbits appearing
at large deformations lead to the shell structure responsible for the
fission isomers known from seventies, which have superdeformed shapes.
As emphasized in Ref.~\citen{stru}, shell structures obtained for the
spheroidal cavity model contain basic features, apart from shifts of
deformed magic numbers due to the spin-orbit potential, similar to
those obtained by the Woods-Saxon potential for heavy nuclei and
metallic clusters, and thus it can be used as a simple model in order
to understand a semiclassical origin of the emergence of regular
oscillating patterns in the coarse-grained quantum spectra at large
deformations.

Remarkably, however, two decade after the publication, to the best of
our knowledge, little exploration of this idea has taken place and
their qualitative argument not fully examined by other researchers,
although the spheroidal cavity model has been used for various
purposes\cite{arvieu1,arvieu2,arvieu3}.  A paper most relevant to
the present paper is that of Frisk\cite{frisk} who used the
periodic-orbit theory and the same cavity model mainly to clarify the
origin of the prolate-oblate asymmetry at normal
deformations.  Although he briefly discussed also the case of large
deformations, the importance of 3D orbits was not mentioned.

In this paper, we identify most important periodic orbits that
determine major pattern of the oscillating level density at large
deformations, including prolate superdeformations, prolate
hyperdeformations and oblate superdeformations.  For this purpose we
make full use of the Fourier transformation method.  As briefly
reviewed in the text, by virtue of the scaling property of the cavity
model, Fourier transforms of quantum spectra exhibit peaks at lengths
of classical periodic orbits, enabling us to precisely identify
important periodic orbits contributing to the shell structure.  This
method has been well known\cite{balian,BB}, but not used for the
present subject.

Classical periodic orbits in the spheroidal cavity and their
bifurcations with variation of the axis ratio have been thoroughly
studied by Nishioka et al.\cite{nishi1,nishi2} This paper may be
regarded as a continuation of their work in the sense that we
investigate quantum manifestations of these periodic orbits and of
their bifurcations.  (Actually, this was the intention also of the
work by Nishioka et al.\cite{nishi1,nishi2})

We present in section~\ref{sec:gosc} oscillating parts of smoothed
level densities as function of deformation parameter of the cavity.
The Fourier transformation method is recapitulated in
section~\ref{sec:fourier}.  Periodic orbits and their bifurcations in
the spheroidal cavity are briefly reviewed in
section~\ref{sec:bifurcation}.  Results of semiclassical analysis of
shell structures are presented in sections~\ref{sec:super},
\ref{sec:hyper} and \ref{sec:oblate} for prolate superdeformations, prolate
hyperdeformations and oblate superdeformations, respectively, and
conclusions are given in section~\ref{sec:conclusion}.

A part of this work was previously reported in conference
proceedings.\cite{conf}

\section{Oscillating level density}
\label{sec:gosc}

We solve the Schr\"odinger equation for single-particle motions in the
spheroidal cavity under the Dirichlet boundary condition.  As is well
known, the spheroidal cavity is integrable and separable by the
spheroidal coordinate system, so that these coordinates are frequently
used for solving the Schr\"odinger equation.  We have, however,
adopted a spherical-wave decomposition method\cite{pal} for this
purpose.  The reason is merely that we had constructed a computer
program based on the latter method for the purpose of efficiently
calculating a large number of eigenvalues as function of deformation
parameters for cavities of general axially symmetric shapes\cite{SAM}.
In this method, wave functions are expanded in terms of spherical
Bessel functions (for radial coordinate) and associated Legendre
functions (for polar angle coordinate), and expansion coefficients are
determined so as to fulfill the boundary conditions (See
Ref.~\citen{SAM,pal} for technical details).

\begin{figure}
\epsfxsize=.72\textwidth
\centerline{\epsffile{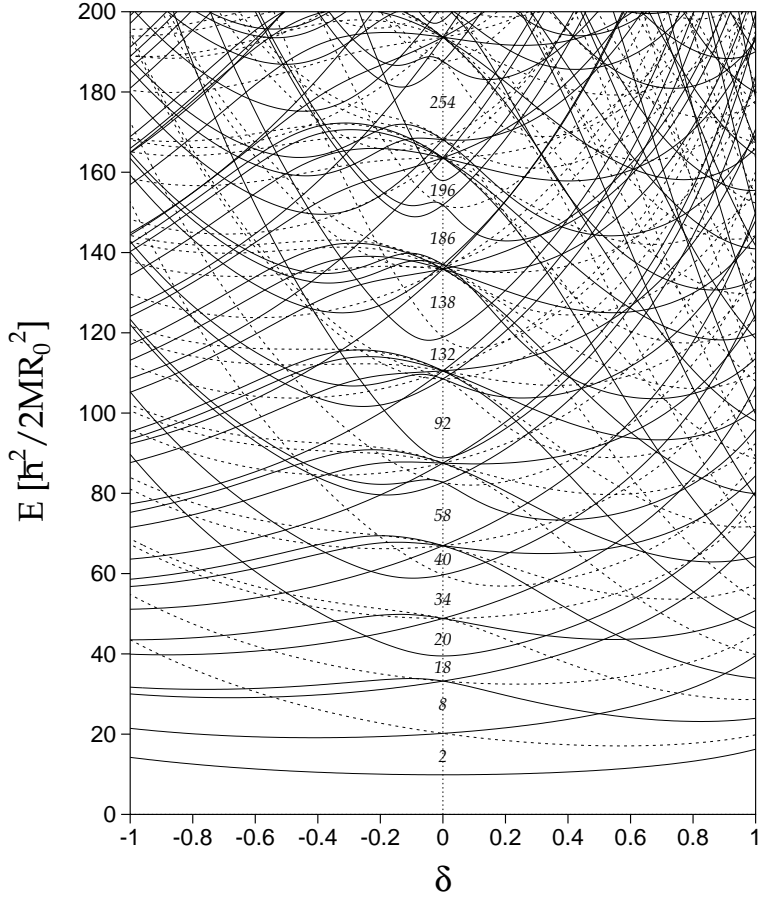}}
\caption{\label{fig:diagram}
Single-particle energy diagram for spheroidal cavity, plotted as
function of deformation parameter $\delta$.  Solid and broken lines
represent even- and odd-parity levels.  The energy is measured in unit
of $\hbar^2/2MR_0^2, M$ and $R_0$ being the mass of a particle and the
radius in the spherical limit, respectively.  Spin degeneracy factor 2 
is taken into account in magic numbers of the spherical limit.}
\end{figure}
The single-particle energy diagram obtained in this way is shown in
Fig.~\ref{fig:diagram} as function of deformation parameter $\delta$,
which is related to the axis ratio $\eta\equiv a/b$ by
$\delta=3(\eta-1)/(2\eta +1)$ in the prolate case and by
$\delta=-3(\eta-1)/(\eta +2)$ in the oblate case, $a$ and $b$ denoting
lengths of the major and the minor axes, respectively, The
volume-conservation condition is imposed so that $ab^2=R_0^3$ in the
prolate case and $a^2b=R_0^3$ in the oblate case, $R_0$ being the
radius in the spherical limit.
\begin{figure}
\epsfxsize=.8\textwidth
\centerline{\epsffile{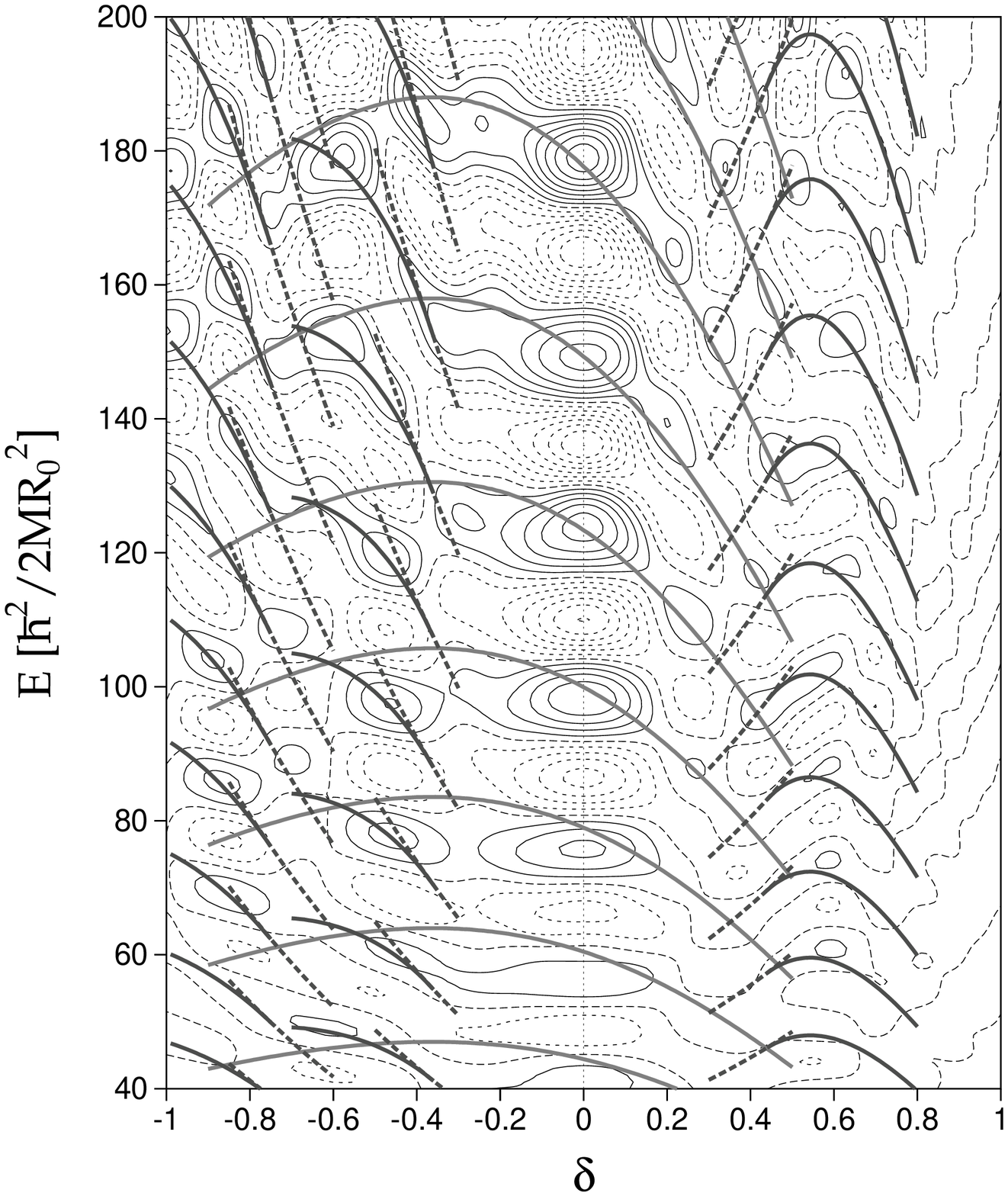}}
\caption{\label{fig:cacmap}
Oscillating part of the smoothed level density displayed as function
of energy and deformation parameter $\delta$.  The unit of energy is
the same as in Fig.~\protect\ref{fig:diagram}.  Strutinsky smoothing
is taken with smoothing width parameter $\Delta k=0.5$.
Constant-action lines for important periodic orbits are indicated:
Thick solid lines running through the spherical closed shells are
those for tetragonal orbits in the meridian plane.  Thick broken and
solid lines in the region $\delta=0.3\sim0.8$ are those for five-point
star-shaped orbits in the equatorial plane and for 3D orbits (5:2:1)
bifurcated from them, respectively.  Broken and solid lines in the
region $\delta=-0.3\sim-0.7$ are those for double repetitions of
linear orbits along the minor axis and for butterfly-shaped planar
orbits (4:1:1) bifurcated from them, respectively.  Likewise, broken
and solid lines in the region $\delta=-0.6\sim-1$ are those for triple
repetitions of linear orbits along the minor axis and for planar
orbits (6:1:1) bifurcated from them, respectively.}
\end{figure}

\begin{figure}
\epsfxsize=\textwidth
\centerline{\epsffile{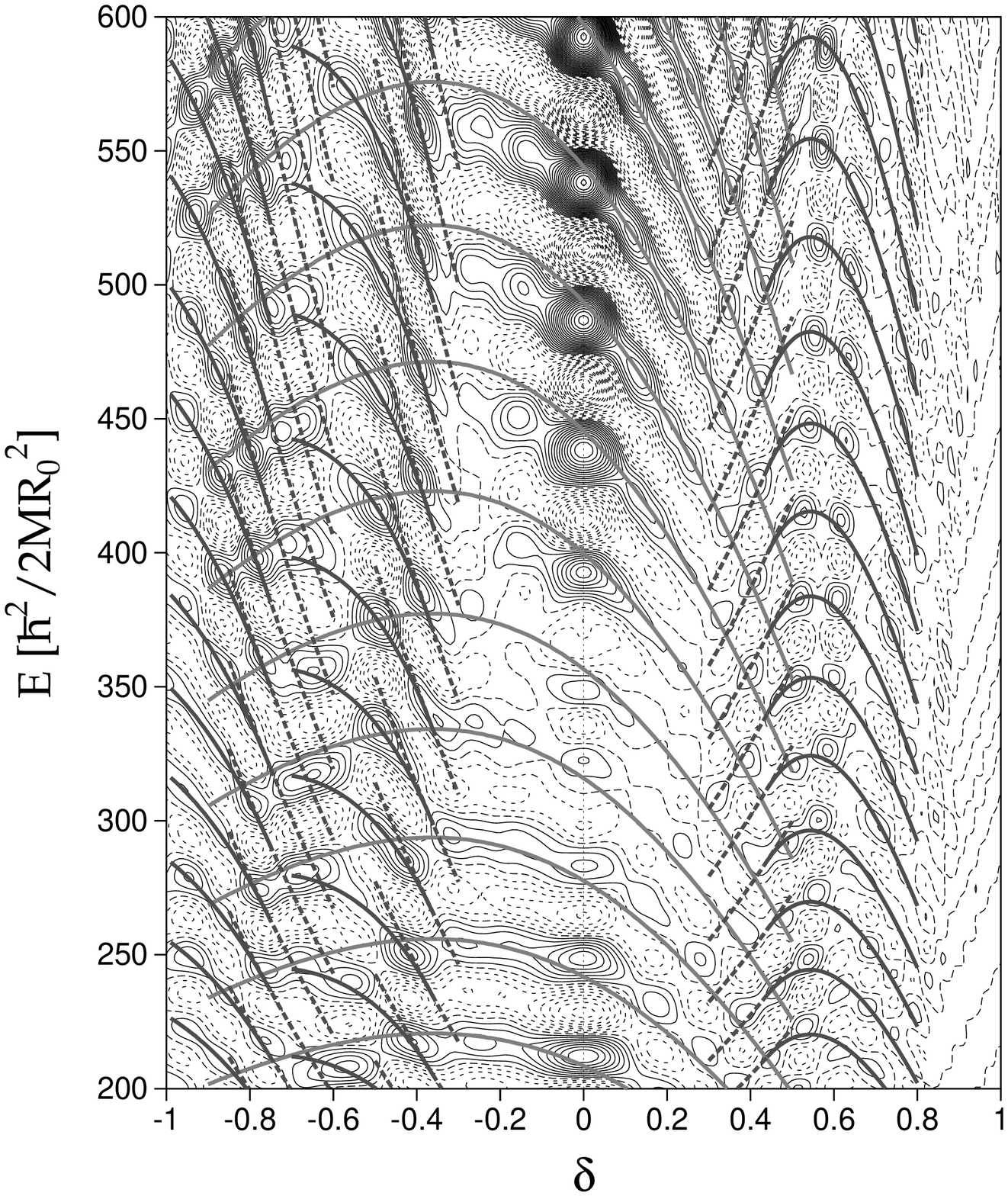}}
\caption{\label{fig:cacmap2}
Same as Fig.~\protect\ref{fig:cacmap} but for higher energy region.
Much clearer classical-quantum correspondences can be seen.}
\end{figure}

Figures~\ref{fig:cacmap} and \ref{fig:cacmap2} display the
oscillating part of the smoothed level density in a form of contour
map with respect to energy and deformation parameter, which is
coarse-grained with the Strutinsky smoothing parameter $\Delta k=0.5$.
We see clearly regular patterns consisting of several valley and ridge
structures.  Thick solid and broken lines indicate constant-action
lines for important periodic orbits responsible for these structures,
implication of which will be the major theme in sections
\ref{sec:super}--\ref{sec:oblate}.

\section{Fourier transform}
\label{sec:fourier}

Single-particle equations of motion for the cavity are invariant with
respect to the scaling transformation $(\bbox{x},\bbox{p},t) \to
(\bbox{x},\alpha\bbox{p},\alpha^{-1}t)$.  The action integral $S_r$
for a periodic orbit $r$ is proportional to the length $L_r$
of it,
\begin{equation}
S_r(E=p^2/2M)=\oint_r\bbox{p}\cdot d\bbox{q}=pL_r=\hbar kL_r,
\end{equation}
and the semiclassical trace formula for the level density\cite{gutw}
is written as
\begin{eqnarray}
g(E)&=&\sum_n\delta(E-E_n)=\frac{M}{\hbar^2k}\sum_n\delta(k-k_n) \nonumber\\
&\simeq&\bar{g}(E)+\sum_r A_r(k)\,\cos(kL_r-\pi\mu_r/2),
\label{eq:trace}
\end{eqnarray}
where $\bar{g}(E)$ denotes the smooth part corresponding to the
contribution of {\it zero-length} orbit, $\mu_r$ the Maslov
phase of the periodic orbit $r$.  This scaling property enables
us to make use of the Fourier transformation of the level density with
respect to the wave number $k$.  The Fourier transform $F(L)$ of the
level density $g(E)$ is written as
\begin{eqnarray}
F(L)&=&\int dk\, e^{-ikL} g(E=\hbar^2k^2/2M) \nonumber\\
    &\simeq&\bar{F}(L) + \pi\sum_r e^{-i\pi\mu_r/2}
    A_r(i\partial_L)\,\delta(L-L_r),
\end{eqnarray}
which may be regarded as `length spectrum' exhibiting peaks at lengths
of individual periodic orbits.\cite{balian}  In numerical calculation,
the spectrum is cut off by Gaussian with cut-off wave number
$k_c=1/\Delta L$ as
\begin{eqnarray}
F_{\Delta L}(L)&\equiv&\frac{1}{\sqrt{2\pi}\Delta L}\int dL'\,
e^{-\frac12(\frac{L-L'}{\Delta L})^2}\,F(L')
\nonumber \\
&=&\frac{M}{\hbar^2}\sum_n \frac{1}{k_n}\,e^{-\frac12(k_n/k_c)^2}\,
  e^{-ik_nL}  \label{eq:fourier1} \\
&\simeq&\bar{F}_{\Delta L}(L) + \pi \sum_r e^{-i\pi\mu_r/2}
  A_r(i\partial_L)\,\frac{1}{\sqrt{2\pi}\Delta L}\,
  e^{-\frac12(\frac{L-L_r}{\Delta L})^2}.
\label{eq:fourier2}
\end{eqnarray}

\section{Periodic-orbit bifurcations}
\label{sec:bifurcation}

In this section, we recapitulate classical periodic orbits in the
spheroidal cavity according to Nishioka et al.\cite{nishi1,nishi2} and
Strutinsky et al.\cite{stru} We focus our attention on those having
short periods.

As is well known, only linear and planar orbits exist in the spherical
cavity.  When spheroidal deformations set in, the linear(diameter)
orbits bifurcate into those along the major axis and along the minor
axis.  Likewise, the planar orbits bifurcate into orbits in the
meridian plane and those in the equatorial plane.  Since the
spheroidal cavity is integrable, all classical orbits lie on a 3D
torus and, in the case of prolate spheroid, periodic orbits are
characterized by three positive integers $(p,t,q)$, which represent
numbers of vibrations or rotations with respect to three spheroidal
coordinates.  They are denoted as $(n_{\epsilon},n_{\phi},n_{\xi})$
and $(n_v,n_{\phi},n_u)$ in Refs.~\citen{nishi1,nishi2} and
\citen{stru}, respectively.  When the axis ratio $\eta$ of the prolate
spheroid increases, hyperbolic orbits in the meridian plane and
three-dimensional orbits are successively born through bifurcations of
linear and planar orbits in the equatorial plane.  Bifurcations occur
when the following condition is satisfied:
\begin{equation}
\eta \equiv \frac ab = \frac{\sin(\pi t/p)}{\sin(\pi q/p)}.
\end{equation}

As we shall see in succeeding sections, most important orbits for
superdeformed shapes (axis ratio about 2:1) are 3D ones having ratio
$\mbox{($p$:$t$:$q$)}=\mbox{($p$:2:1)}$ with $p=5,6,7,\ldots$ etc.
They bifurcate from planar orbits that turns twice ($t=2$) about the
symmetry axis.  Likewise, planar orbits with (4:2:1) bifurcate from
linear orbits that repeat twice along the minor axis.  These new-born
orbits resemble the Lissajous figures of the superdeformed harmonic
oscillator with frequency ratio
$\mbox{$\omega_\perp$:$\omega_z$}=\mbox{2:1}$.  Every bifurcated orbits
form continuous families of degeneracy two, which means that we need
two parameters to specify a single orbit among continuous set of
orbits belonging to a family having a common value of action
integral (or equivalently, length).

For prolate hyperdeformed shapes (axis ratio about 3:1), bifurcations
from linear and planar orbits that turns three times ($t=3$) about the
symmetry axis are important.  The new-born orbits are hyperbolic
orbits in the meridian plane (6:3:1) and 3D orbits ($p$:3:1) with
$p=7,8,9,\ldots$ etc.

In the case of oblate spheroidal cavities, periodic orbits are
classified in Ref.~\citen{nishi2} into two modes; whispering-gallery
(W) mode and bouncing-ball (B) mode.  Systematics of periodic-orbit
bifurcations in the W-mode is similar to the prolate case and can be
treated just by exchanging the role of $t$ and $q$.  On the other
hand, B-mode orbits are successively created through bifurcations of
multiple repetitions of linear orbits along the minor axis, when the
following condition is satisfied:
\begin{equation}
\eta \equiv \frac ab = \frac 1{\sin(\pi t/p)}.
\end{equation}
Bifurcations of this kind do not depend on $q$, so that, for instance,
planar orbits (4:1:1) are created simultaneously with two families of
3D orbits (4:1:3/2) and (4:1:2).  (For B-mode orbits, half integer
values of $q$ are allowed as well as integers, due to different
definition of integration range for the action integral related to
$q$; see Ref.~\citen{nishi2})

\begin{figure}
\epsfxsize=.8\textwidth
\centerline{\epsffile{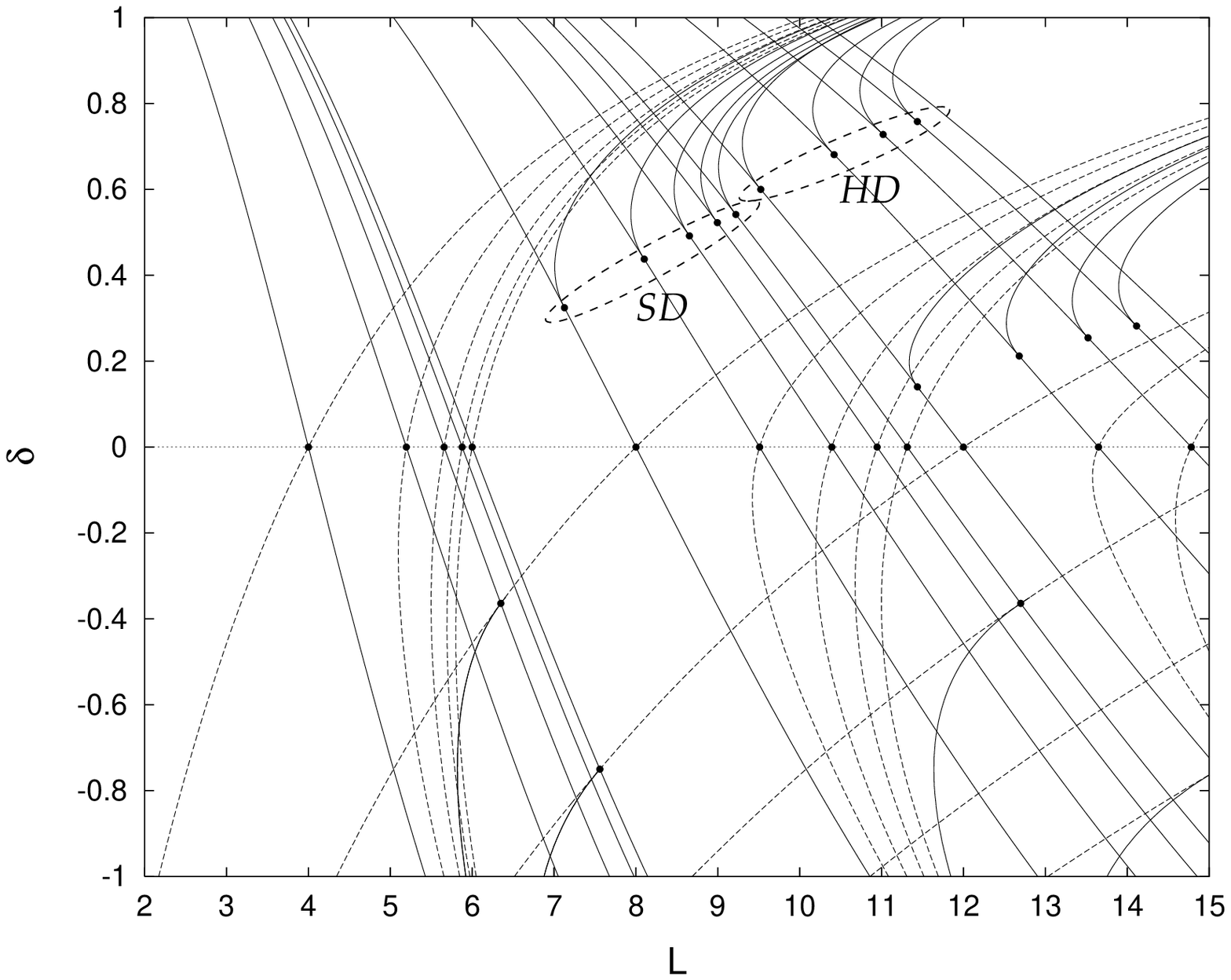}}
\caption{\label{fig:length}
Variations of lengths of periodic orbits specified by ($p$:$t$:$q$) in
the spheroidal cavity with respect to deformation parameter
$\delta$.  Only those for short orbits to be discussed in sections
\ref{sec:super}--\ref{sec:oblate} are displayed.  For a more complete
diagram, see Nishioka et al.\protect\cite{nishi1,nishi2}}
\end{figure}

Bifurcation points and variations of lengths with deformation are
displayed for some short periodic orbits in Table~\ref{tab:orbit} and
Fig.~\ref{fig:length}.

\begin{table}[t]
\catcode`?=\active \def?{\phantom{0}}
\caption{\label{tab:orbit}
Bifurcation points of periodic orbits specified by ($p$:$t$:$q$) in
the spheroidal cavity.  Only those for short orbits to be discussed in
sections~\protect\ref{sec:super}--\protect\ref{sec:oblate} are displayed.}
\begin{center}
\begin{tabular}{cccc}
orbit ($p$:$t$:$q$) & axis ratio $(a/b)$ & deformation $\delta$
& orbit length in $R_0$ \\
\hline
(4:2:1) & 1.414 &  0.325 & ?7.127 \\
(5:2:1) & 1.618 &  0.438 & ?8.101 \\
(6:2:1) & 1.732 &  0.492 & ?8.654 \\
(7:2:1) & 1.802 &  0.523 & ?8.995 \\
(8:2:1) & 1.848 &  0.542 & ?9.220 \\
\hline
(6:3:1) & 2.0?? &  0.6?? & ?9.524 \\
(7:3:1) & 2.247 &  0.681 & 10.421 \\
(8:3:1) & 2.414 &  0.728 & 11.011 \\
(9:3:1) & 2.532 &  0.758 & 11.437 \\
\hline
(4:1:1) & 1.414 & \llap{$-$}0.364 & ?6.350 \\
(6:1:1) & 2.0?? & \llap{$-$}0.75? & ?7.560
\end{tabular}
\end{center}
\end{table}

\section{Shell structure and constant-action line}

\begin{figure}
\epsfxsize=.6\textwidth
\centerline{\epsffile{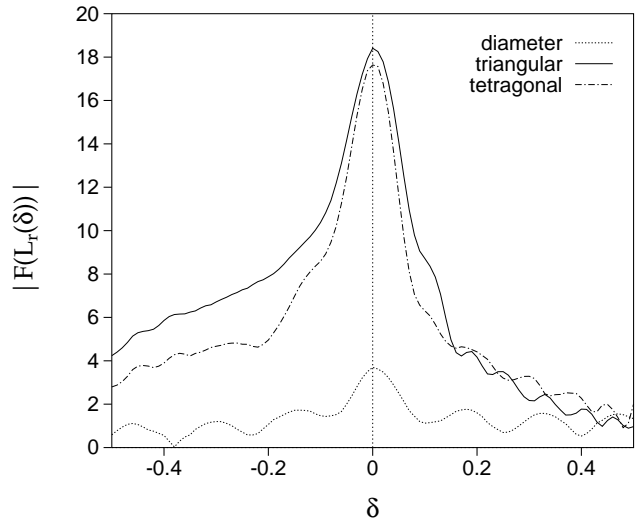}}
\caption{\label{fig:fpeek_nd}
Absolute values of the Fourier amplitudes defined in
Eq.~(\protect\ref{eq:fourier1}), at lengths $L=L_r$ of some short
meridian-plane orbits, plotted as functions of deformation
parameter $\delta$.}
\end{figure}

Using the trace formula, we can extract informations about classical
periodic orbits from the Fourier transforms of the level density.  In
this section we discuss another way of using the trace formula, i.e.,
the constant-action line analysis.\cite{stru}  As stated in the above
section (see Eq.~(\ref{eq:trace})), the quantum level density can be
represented as the summation over periodic orbits.  If a few orbits
having nearly the same action integral dominate in the sum, it is
expected that valleys in the contour map of the oscillating part of
the smoothed level density versus energy $E$ and deformation
$\delta$ are characterized by constant-action lines
$S(E,\delta)=\mbox{const.}$ for those dominant orbits.  The equation
for such lines is $kL_r-\pi\mu_r/2=(2n+1)\pi$, namely,
\begin{equation}
E(\delta)=\frac{1}{2M}\left(\frac{2\pi\hbar(n+1/2+\mu_r/4)}{L(\delta)}
\right)^2, \quad (n=0,1,2,\cdots).
\label{eq:caline}
\end{equation}
As an example, let us examine the shell structure at normal deformations
$|\delta|\lsim0.3$.  In this region, triangular and tetragonal orbits
in the meridian plane give dominant contributions to the
level density.
This fact was first pointed out by Strutinsky et al.~\cite{stru}
(Although the triangular orbits were overlooked there, actions of the
two families of orbits scale in the same way as function of
deformation so that their argument was correct in essence.)  The
Fourier amplitudes at lengths of some meridian-plane orbits are
plotted in Fig.~\ref{fig:fpeek_nd} as function of deformation
parameter $\delta$.
We see that the meridian-plane orbits are important for small
$\delta$ and their
contributions decline with increasing $|\delta|$.
In Figs.~\ref{fig:cacmap} and \ref{fig:cacmap2}, constant-action lines
(\ref{eq:caline}) for the tetragonal orbits in the meridian plane are
indicated. 
The period of the shell oscillation is mainly determined by the
tetragonal orbits, and the valley structure of 
the level density at normal
deformation is nicely explained by the constant-action lines of them.
We also notice that the intensity of spherical shell structure is
weakened at $E\sim 300$
and the phase of valley is shifted from that of
the constant-action lines for
$E\simeq 250\mbox{--}350$.  This is due to the supershell
effect associated with the interference of the triangular and tetragonal
orbits.\cite{balian,NHM}

In this way, we can analyze the properties of the shell structure through
classical periodic orbits.  In the following sections, we shall utilize
these techniques in order to identify dominant classical periodic orbits
that characterize the shell structures in superdeformed shapes.

\section{Prolate superdeformations}
\label{sec:super}

\begin{figure}
\centerline{\epsfxsize=.75\textwidth\epsffile{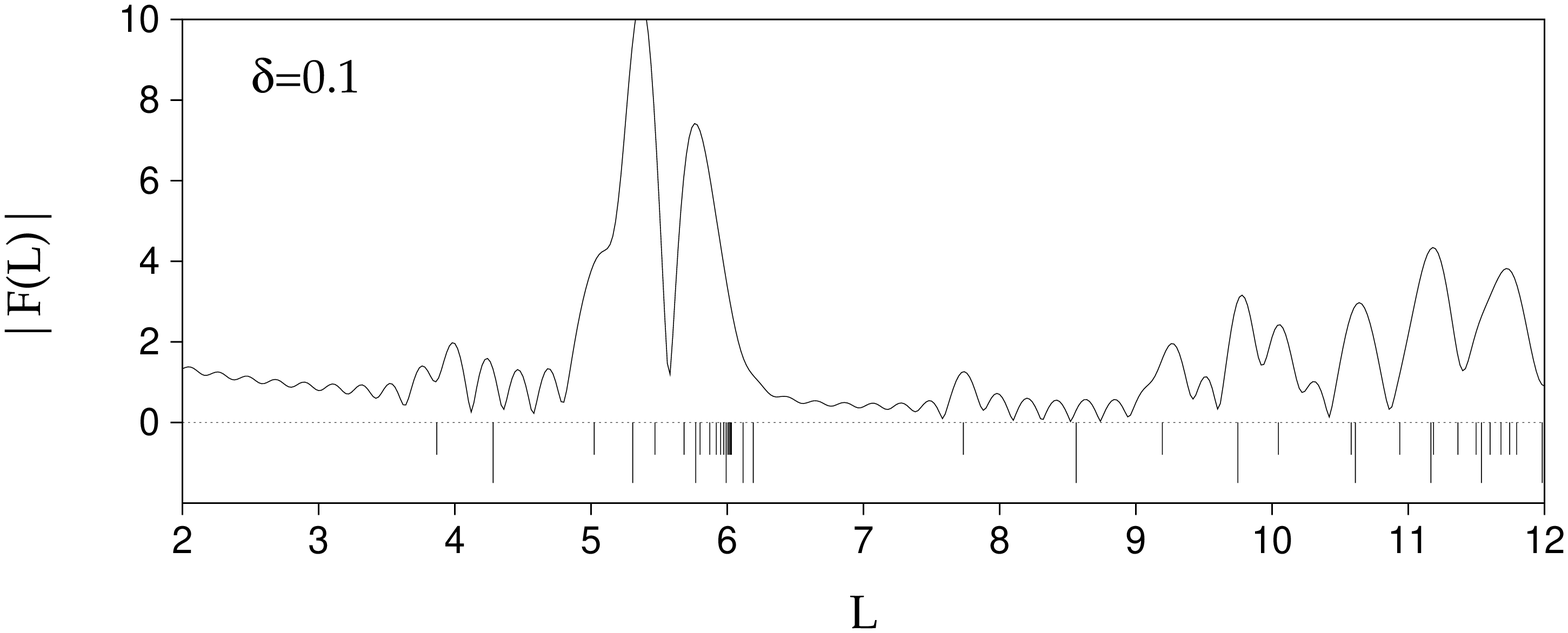}}
\centerline{\epsfxsize=.75\textwidth\epsffile{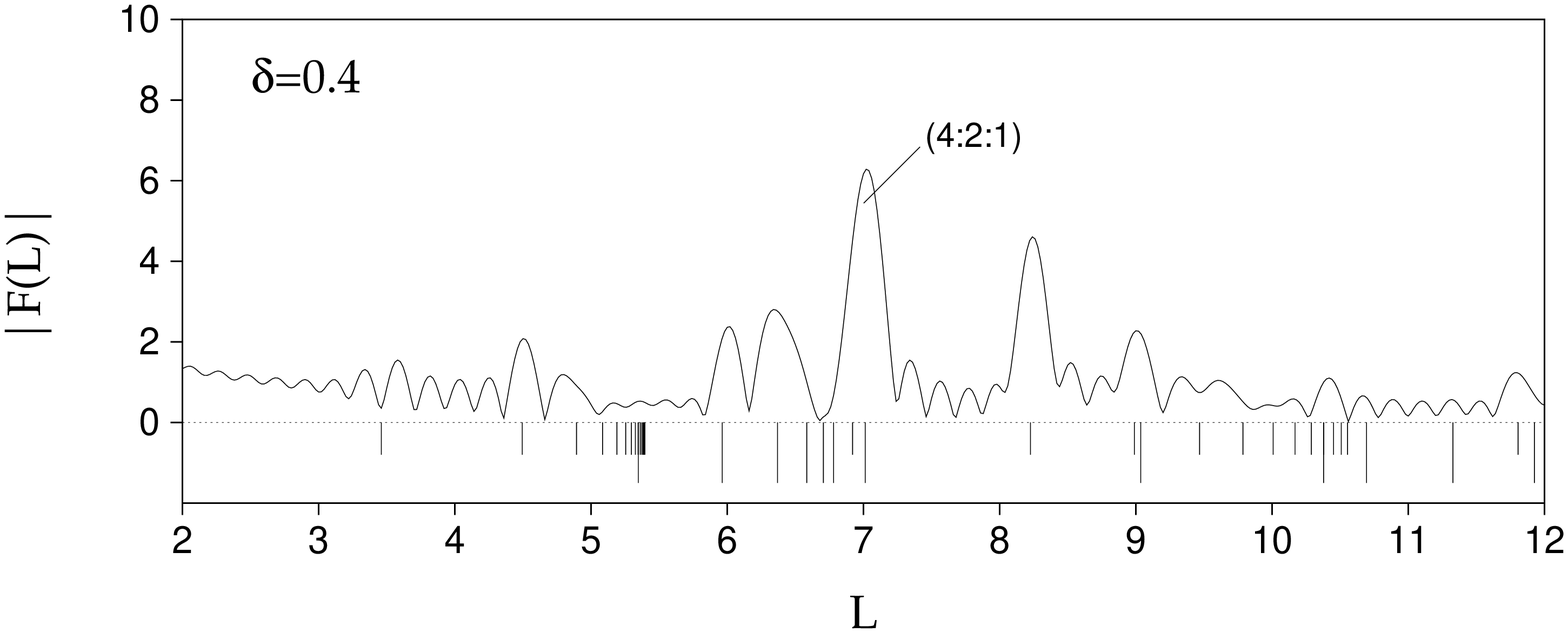}}
\centerline{\epsfxsize=.75\textwidth\epsffile{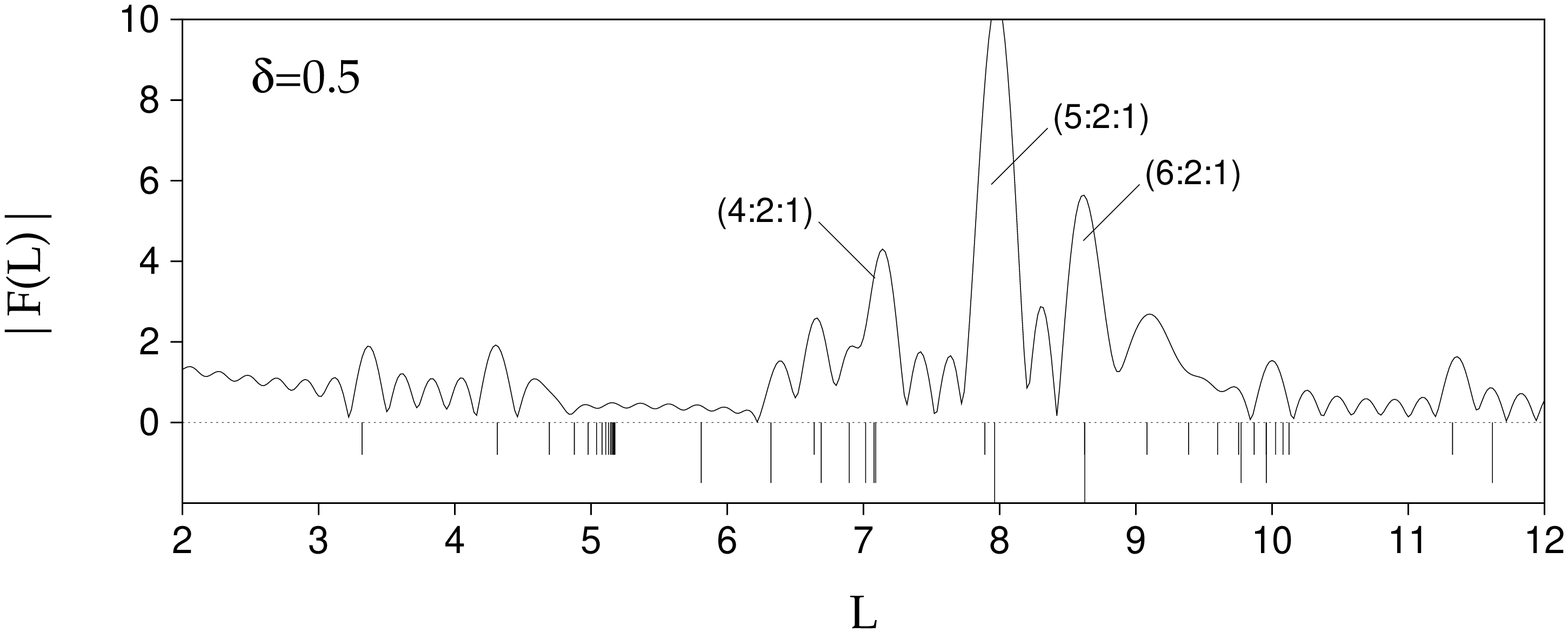}}
\centerline{\epsfxsize=.75\textwidth\epsffile{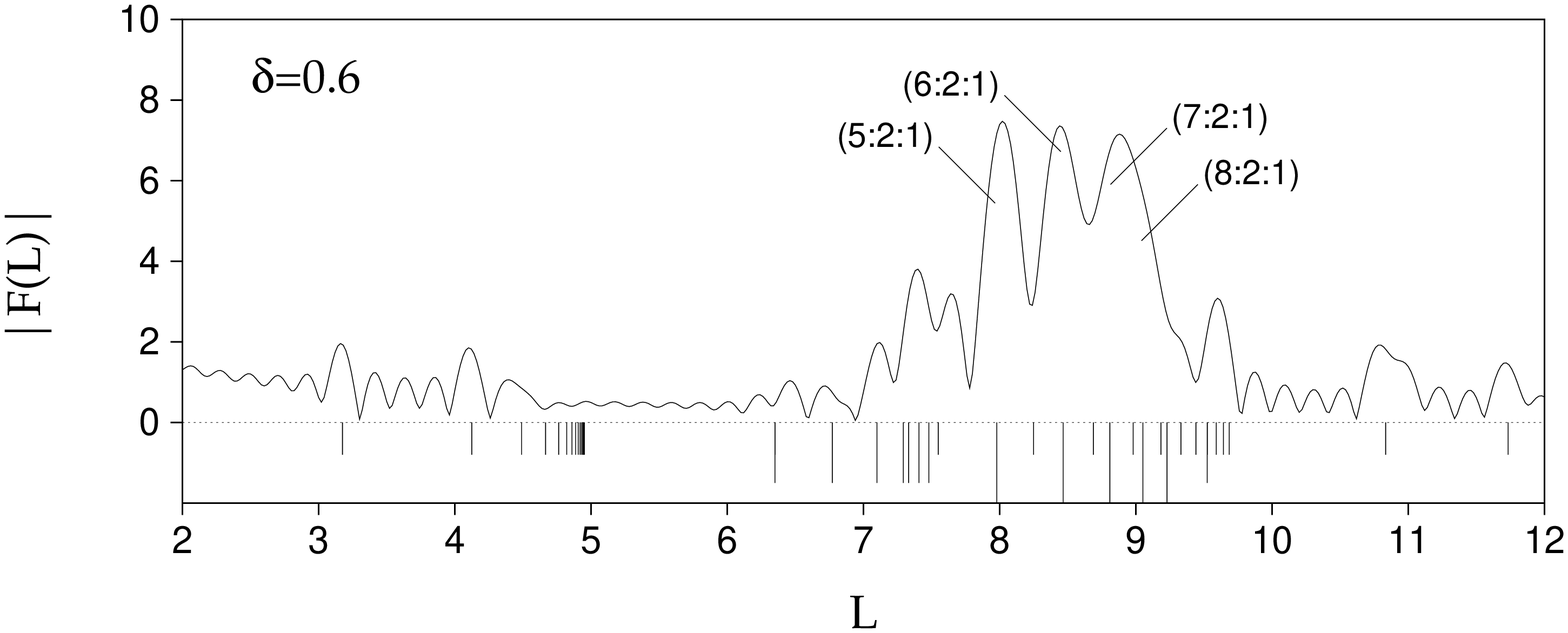}}
\caption{\label{fig:ftl-sd}
Length spectra, i.e., Fourier transforms of quantum level densities,
for spheroidal cavities with deformation parameter $\delta=0.1, 0.4,
0.5$ and $0.6$.  Cut-off wave number $k_c=\protect\sqrt{600}$ is used
in Eq.~(\protect\ref{eq:fourier1}).  In the bottoms of every figures,
lengths of classical periodic orbits are indicated by vertical lines.
Long, middle and short vertical lines are used for 3D orbits, planar
orbits in the meridian and the equatorial planes, respectively.}
\end{figure}

Figure~\ref{fig:ftl-sd} displays Fourier transforms of quantum spectra
for prolate spheroidal cavities with deformation parameter
$\delta=0.1\sim 0.6$.
At normal deformations with $\delta=0.1$, as mentioned in the previous
section, we notice peaks associated with the triangular and tetragonal
orbits in the meridian plane.
With increasing deformation, bifurcations of linear and planar orbits
in the equatorial plane successively take places.  Thus, the highest
peak at $L\simeq 7$ of the Fourier transform for $\delta=0.4$ is
associated with butterfly-shaped planar orbits with
$\mbox{($p$:$t$:$q$)}=\mbox{(4:2:1)}$, that bifurcate at $\delta\simeq
0.32$ from double repetitions of linear orbits along the minor axis.
For $\delta=0.5$, the prominent
peak at $L\simeq 8$ and $8.6$ correspond to 3D orbits (5:2:1) and
(6:2:1) bifurcated respectively from five-point star-shaped orbits and
double traversals of triangular orbits in the equatorial plane.  With
further increase of $\delta$, same kinds of 3D orbits successively
bifurcate from equatorial-plane orbits.  For $\delta=0.6$ (axis ratio
$\eta=2$), peaks around $L\simeq 9$ are associated with 3D orbits
(7:2:1) and (8:2:1) that are bifurcated from 7-point star-shaped
orbits and double traversals of rectangular orbits in the equatorial
plane.

In Figs.~\ref{fig:cacmap} and \ref{fig:cacmap2}, constant-action lines
for the 3D orbits (5:2:1) are indicated.  Good correspondence is found
between these lines and the valley structure seen in the superdeformed
region with $\delta$ around 0.6.  Thus we can conclude that the
bifurcations of equatorial orbits play essential roles in the
formation of the
superdeformed shell structure, and it is characterized by the 3D
orbits ($p$:2:1).

\begin{figure}
\noindent
\begin{minipage}[b]{.48\textwidth}
\epsfxsize=.9\textwidth\rightline{\epsffile{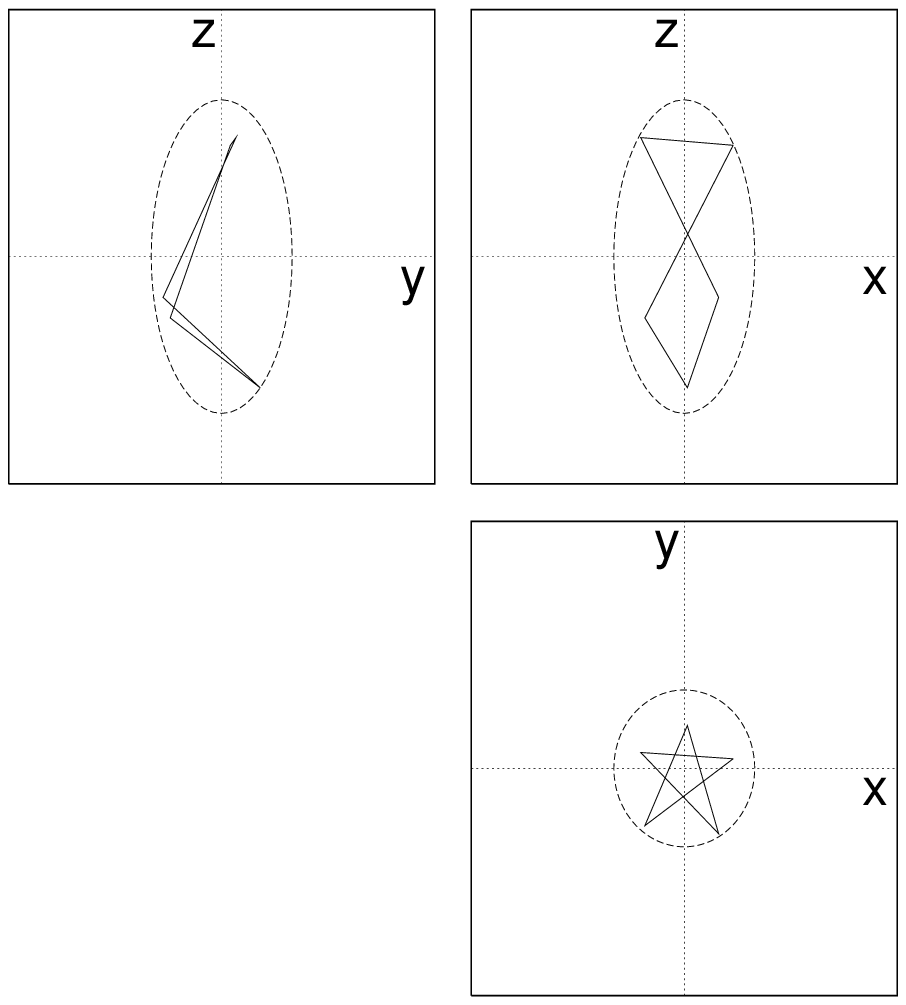}}
\end{minipage}
\hfill
\begin{minipage}[b]{.48\textwidth}
\epsfxsize=.9\textwidth\leftline{\epsffile{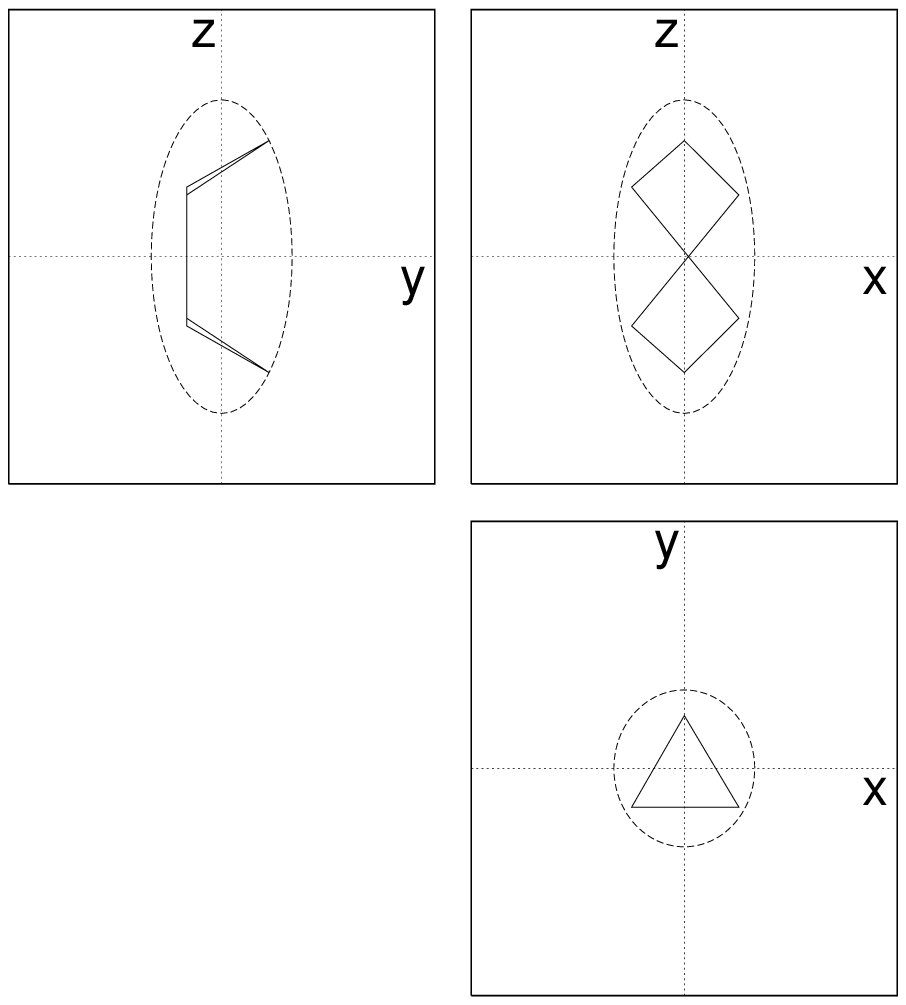}}
\end{minipage}
\caption{\label{fig:orbit-sd}
Three-dimensional orbits (5:2:1) and (6:2:1) in the superdeformed
prolate cavity with deformation $\delta=0.6$ (axis ratio $\eta=2$).
Their projections on the $(x,y)$, $(y,z)$ and $(z,x)$ planes are
displayed.}
\end{figure}

Some of these 3D orbits are displayed in Fig.~\ref{fig:orbit-sd}.
They possess similarities with the figure-eight shaped orbits in the
superdeformed harmonic oscillator with frequency ratio
$\mbox{$\omega_\bot$:$\omega_z$}=\mbox{2:1}$.  An important difference
between the cavity model under consideration and the harmonic
oscillator model should be noted, however: In the former those
periodic orbits exist for all deformation parameters $\delta$ larger
than the bifurcation points, whereas in the latter such orbits appear
only for special deformations corresponding to rational ratios of the
major and the minor axes.

On the other hand, magnitudes of contributions of individual orbits
are found to exhibit remarkable deformation dependence.  Namely,
Fourier peak heights associated with new orbits created by
bifurcations quickly increase with increasing deformation and reach
the maxima.  Then, they start to decline.  This behavior is apparently
seen in Fig.~\ref{fig:ftl-sd}.
\begin{figure}
\noindent
\begin{minipage}[b]{.48\textwidth}
\epsfxsize=\textwidth\epsfbox[80 40 592 460]{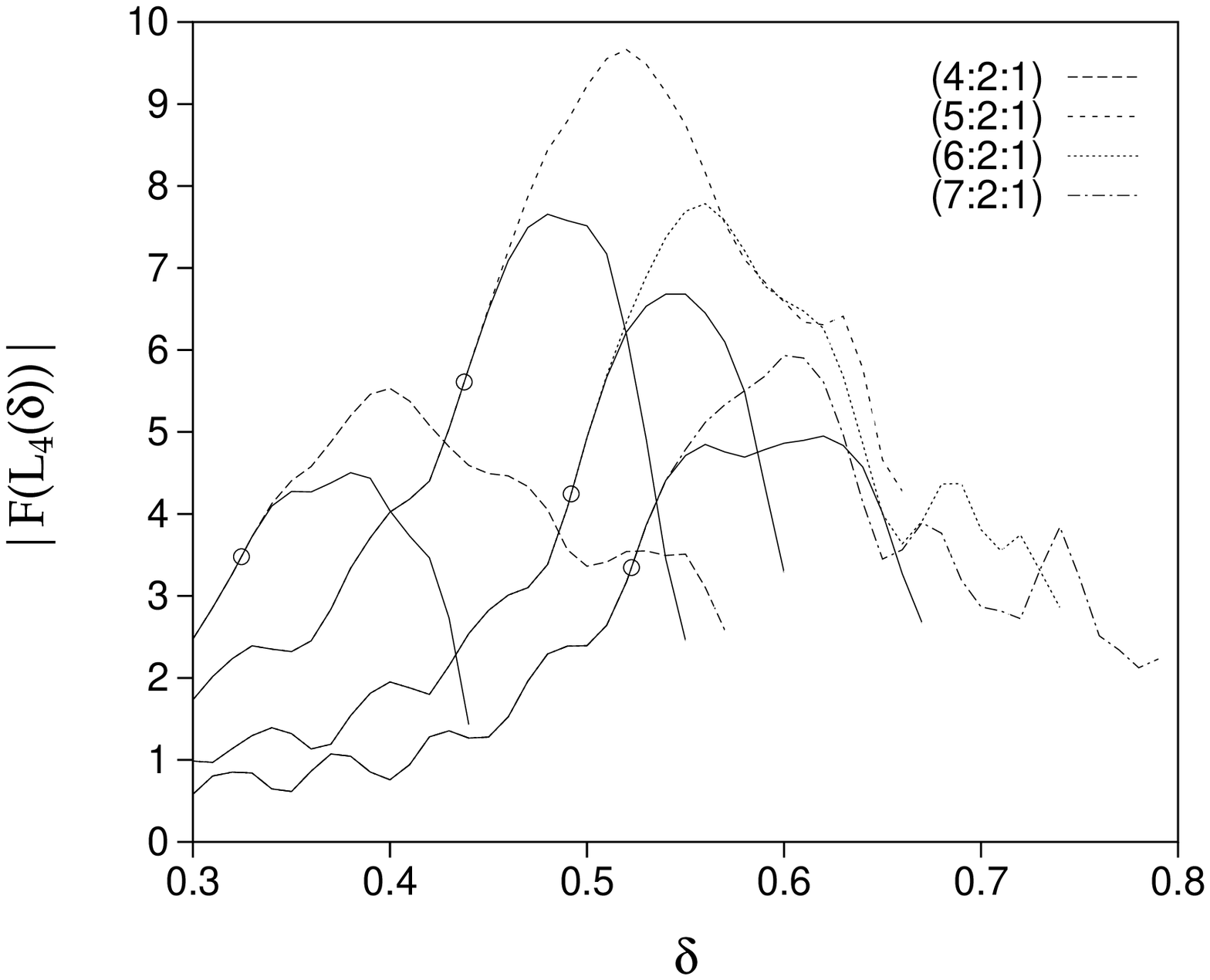}
\end{minipage}
\hfill
\begin{minipage}[b]{.48\textwidth}
\epsfxsize=\textwidth\epsfbox[80 40 592 460]{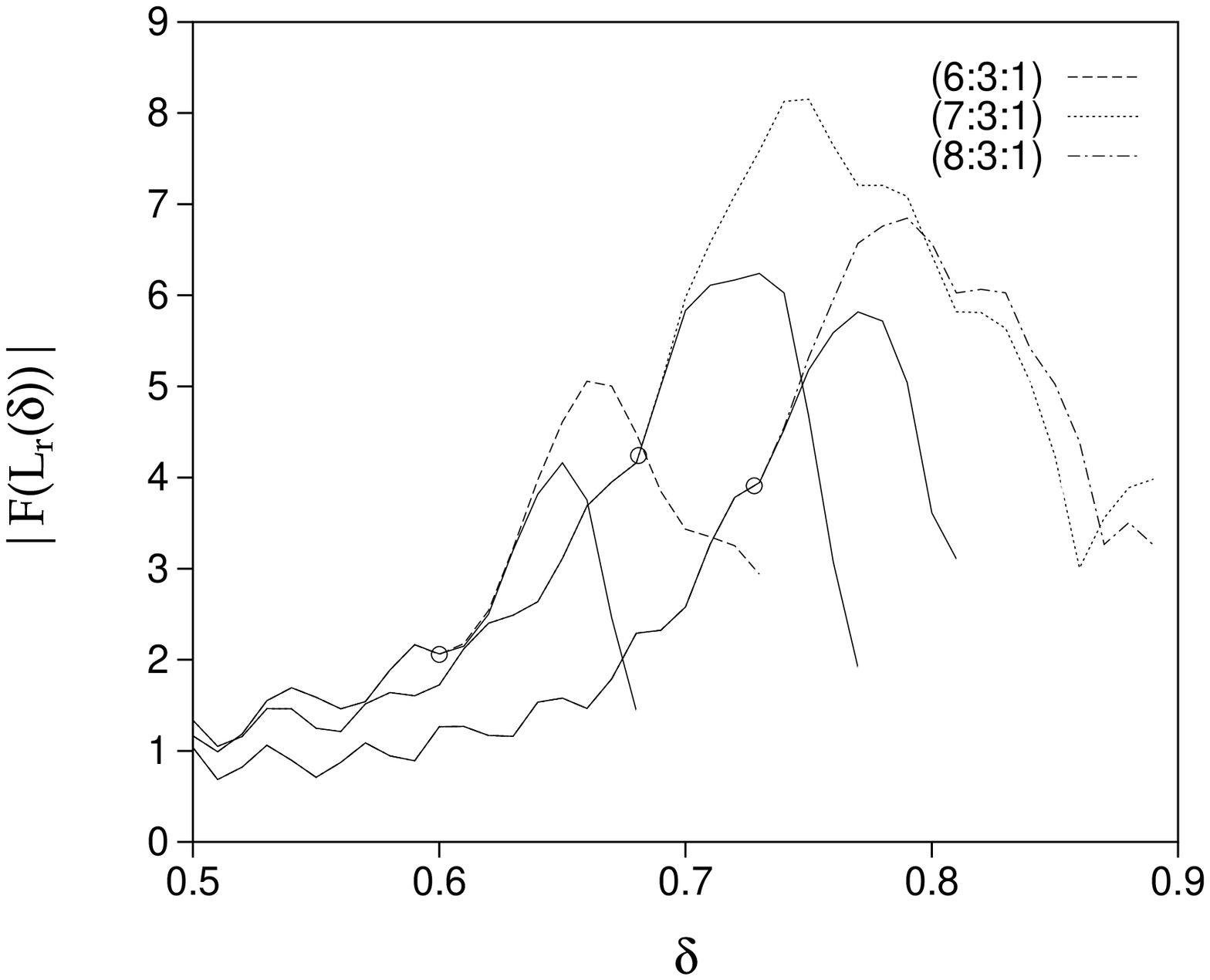}
\end{minipage}
\caption{\label{fig:fpeek_sd}
Same as Fig.~\protect\ref{fig:fpeek_nd} but for meridian-plane
hyperbolic orbits and 3D orbits ($p$:2:1) (left-hand side) and
($p$:3:1) (right-hand side).  Solid lines are those for 
equatorial-plane orbits from which these orbits are bifurcated.}
\end{figure}
Figure~\ref{fig:fpeek_sd} displays deformation dependence of the Fourier
amplitudes $|F(L)|$ defined in Eq.~(\ref{eq:fourier1}) at lengths
$L=L_r$ for some classical periodic orbits, which confirms the
behavior noted above.  This behavior has not been pointed out in the
previous papers.  It should be emphasized that relative magnitudes and
deformation dependence of contributions of individual periodic orbits
found here are significantly different from those roughly estimated in
Ref.~\cite{stru}.

To get deeper understanding on the mechanism of the enhancements
associated with bifurcations noted above, semiclassical
approximation that
goes beyond the stationary phase approximation used in deriving the
trace formula (\ref{eq:trace}) may be required.  Such a semiclassical
theory applicable for three dimensional deformed cavities is not
available at present and remains as a challenging subject for future.

\section{Prolate hyperdeformations}
\label{sec:hyper}

\begin{figure}
\centerline{\epsfxsize=.75\textwidth\epsffile{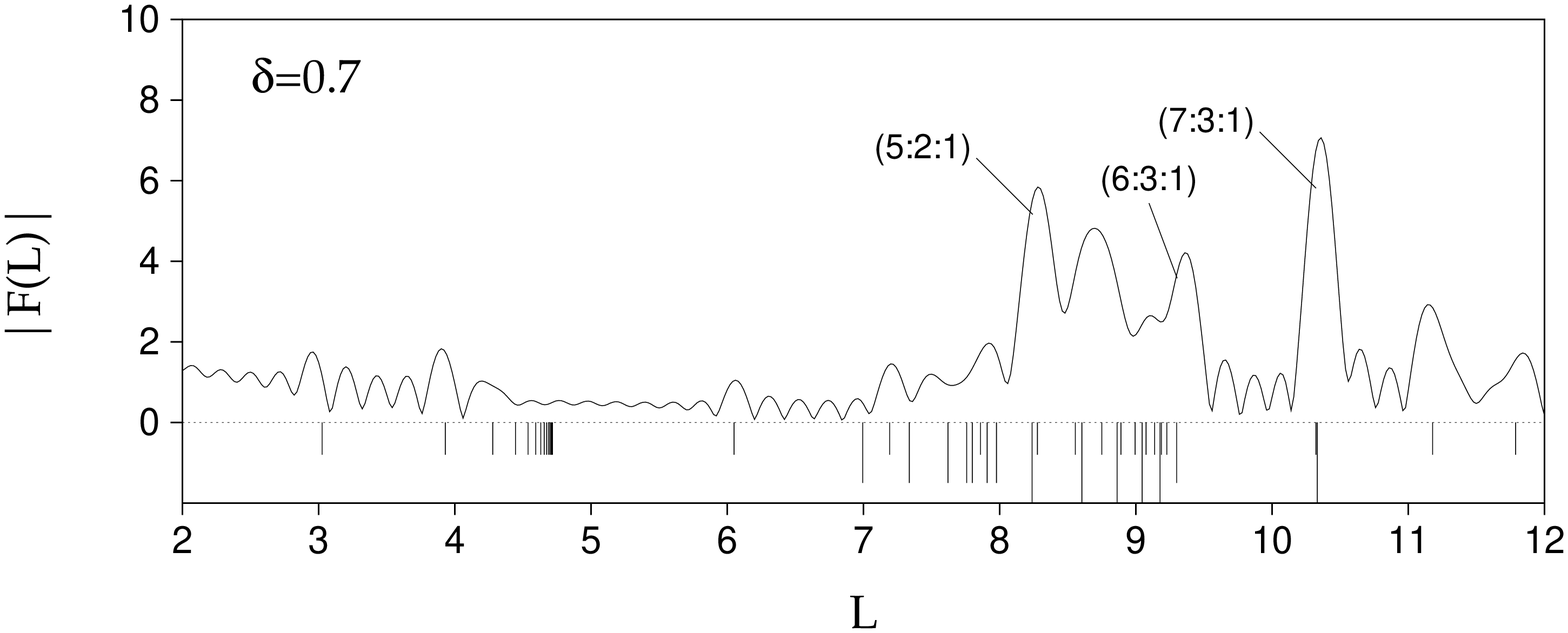}}
\centerline{\epsfxsize=.75\textwidth\epsffile{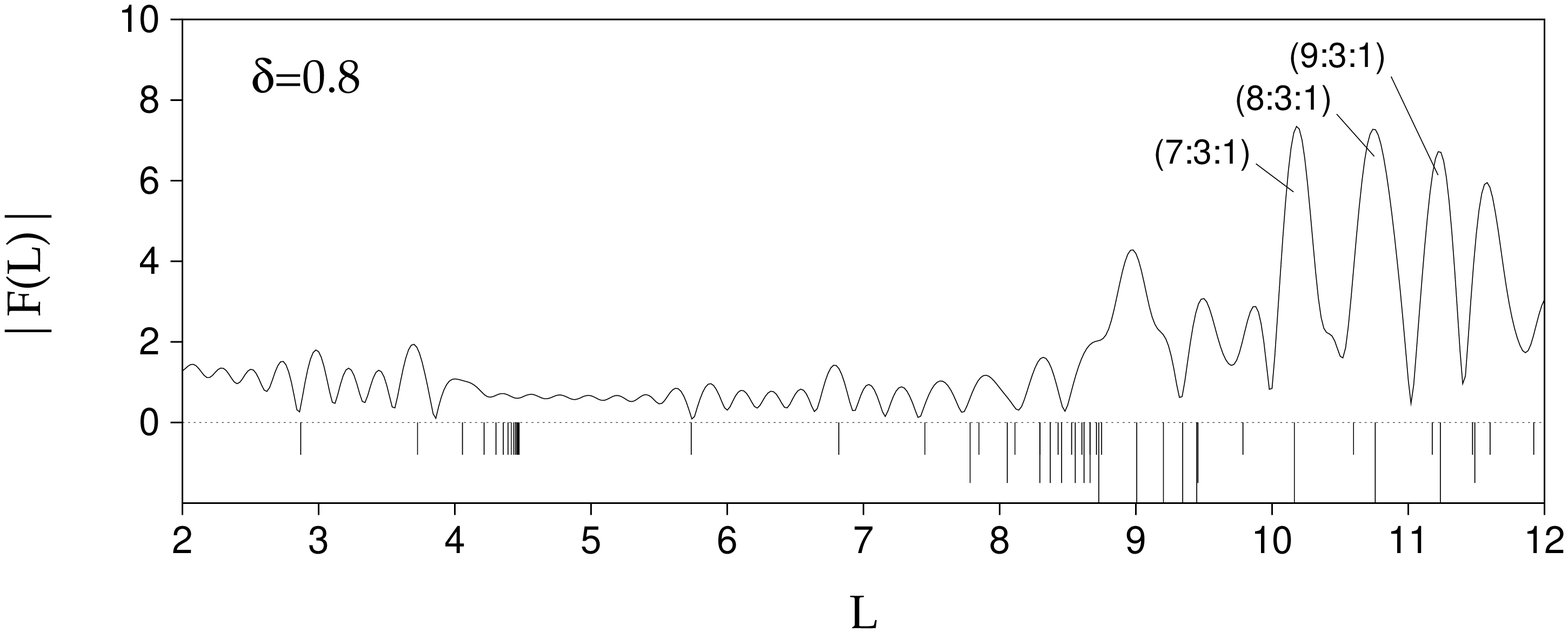}}
\caption{\label{fig:ftl-hd}
Same as Fig.~\protect\ref{fig:ftl-sd} but for $\delta=0.7$ and $0.8$ (axis
ratio $\eta\simeq 2.3$ and $2.7$).}
\end{figure}

\begin{figure}
\noindent
\begin{minipage}[b]{.48\textwidth}
\epsfxsize=.9\textwidth\rightline{\epsffile{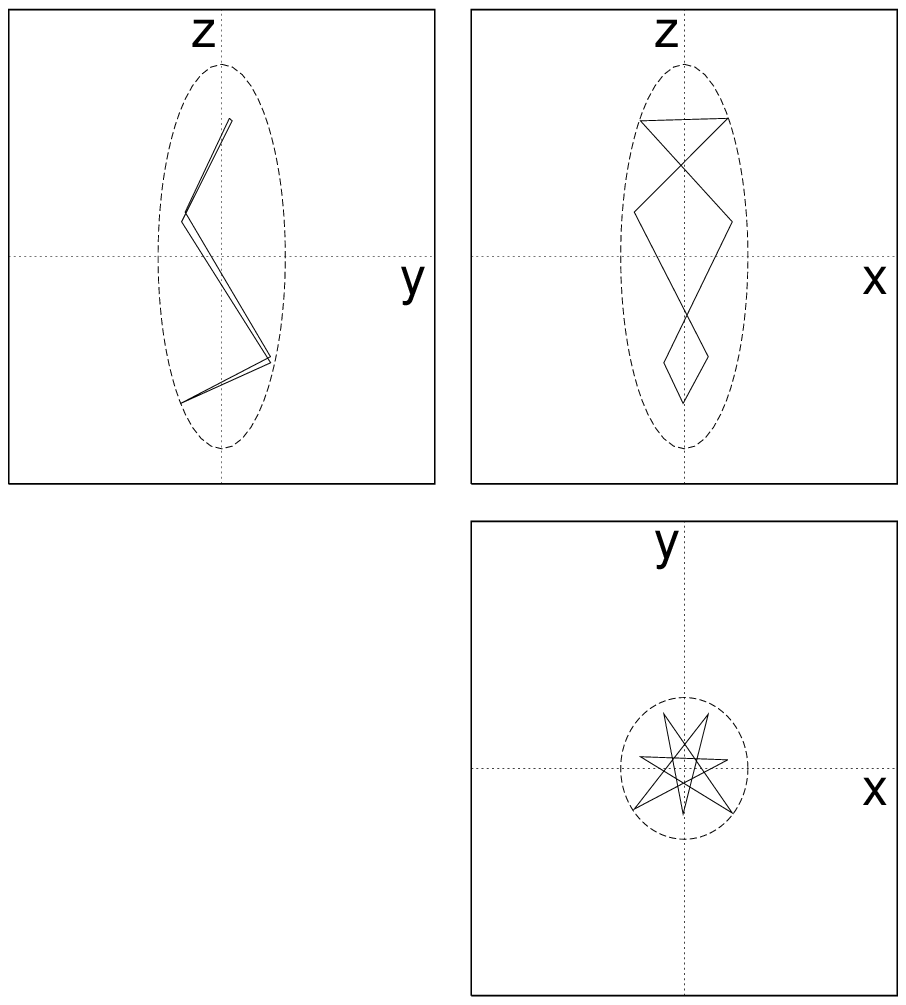}}
\end{minipage}
\hfill
\begin{minipage}[b]{.48\textwidth}
\epsfxsize=.9\textwidth\leftline{\epsffile{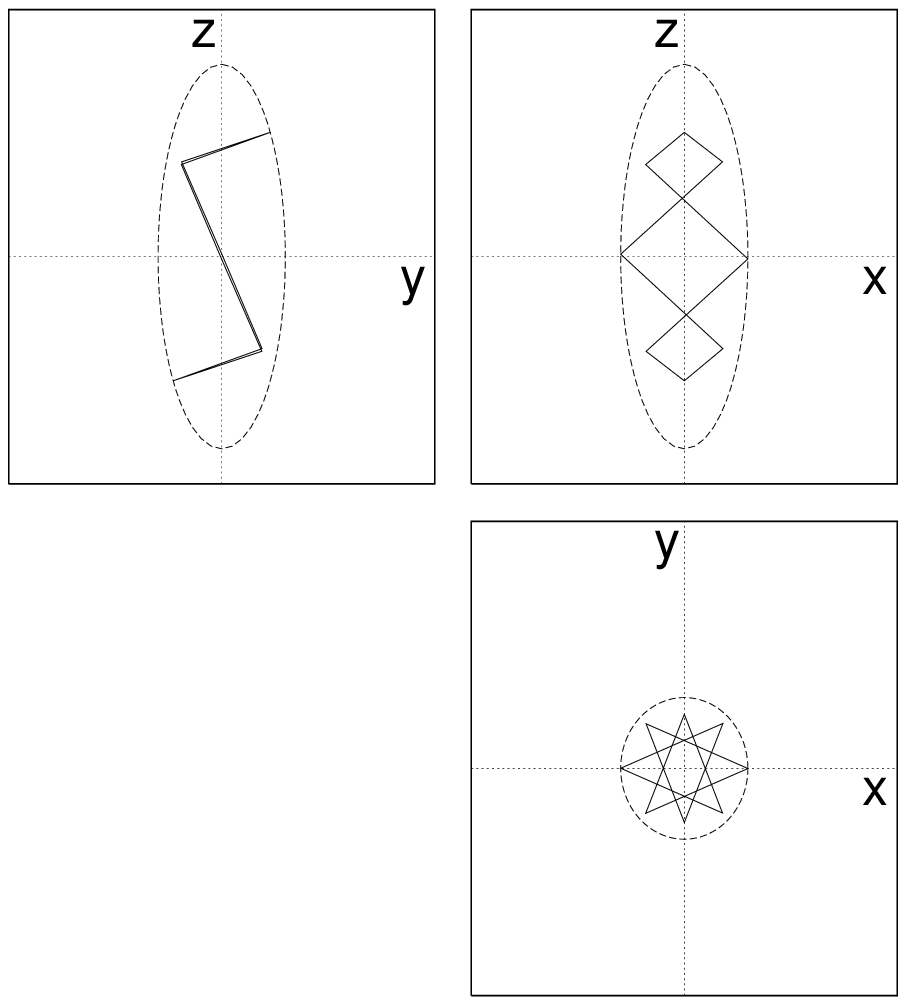}}
\end{minipage}
\caption{\label{fig:orbit-hd}
Same as Fig.~\protect\ref{fig:orbit-sd} but for 3D orbits (7:3:1) and (8:3:1)
in the hyperdeformed prolate cavity with deformation $\delta=0.8$.}
\end{figure}

Figure~\ref{fig:ftl-hd} displays Fourier transforms for $\delta=0.7$
and 0.8.
For $\delta\simeq 0.7$, the peak at $L\simeq 10.3$ is
associated with 3D orbits (7:3:1) bifurcated from 7-point
star-shaped orbits in the equatorial plane.  For $\delta=0.80$, we see
a rise of peak at $L\simeq 10.8$, which is associated with 3D orbits
(8:3:1) bifurcated from 8-point star-shaped orbits in the equatorial
plane.  They are displayed in Fig.~\ref{fig:orbit-hd}.
These 3D orbits resemble with the Lissajous figures of the
hyperdeformed harmonic oscillator with frequency ratio
$\mbox{$\omega_\bot$:$\omega_z$}=\mbox{3:1}$.  In the same manner as
for the 3D orbits responsible for superdeformations, Fourier peak
heights associated with these new-born orbits rapidly increase after
the bifurcations, reach the maxima and then decline with increasing
deformation (see Fig.~\ref{fig:fpeek_sd}).  Thus, also in this case,
bifurcations of equatorial orbits, but of different types ($p$:3:1),
play the major role in the formation of this shell structure.

\section{Oblate superdeformations}
\label{sec:oblate}

\begin{figure}
\centerline{\epsfxsize=.75\textwidth\epsffile{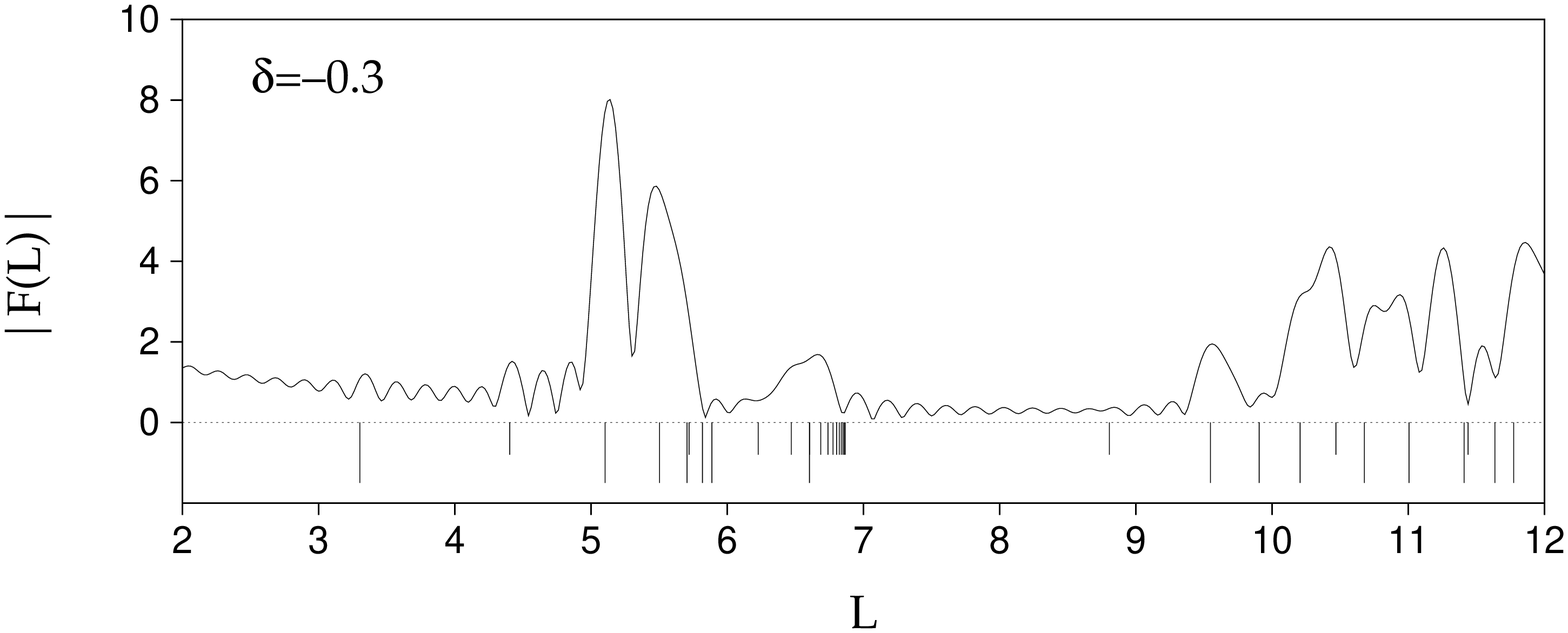}}
\centerline{\epsfxsize=.75\textwidth\epsffile{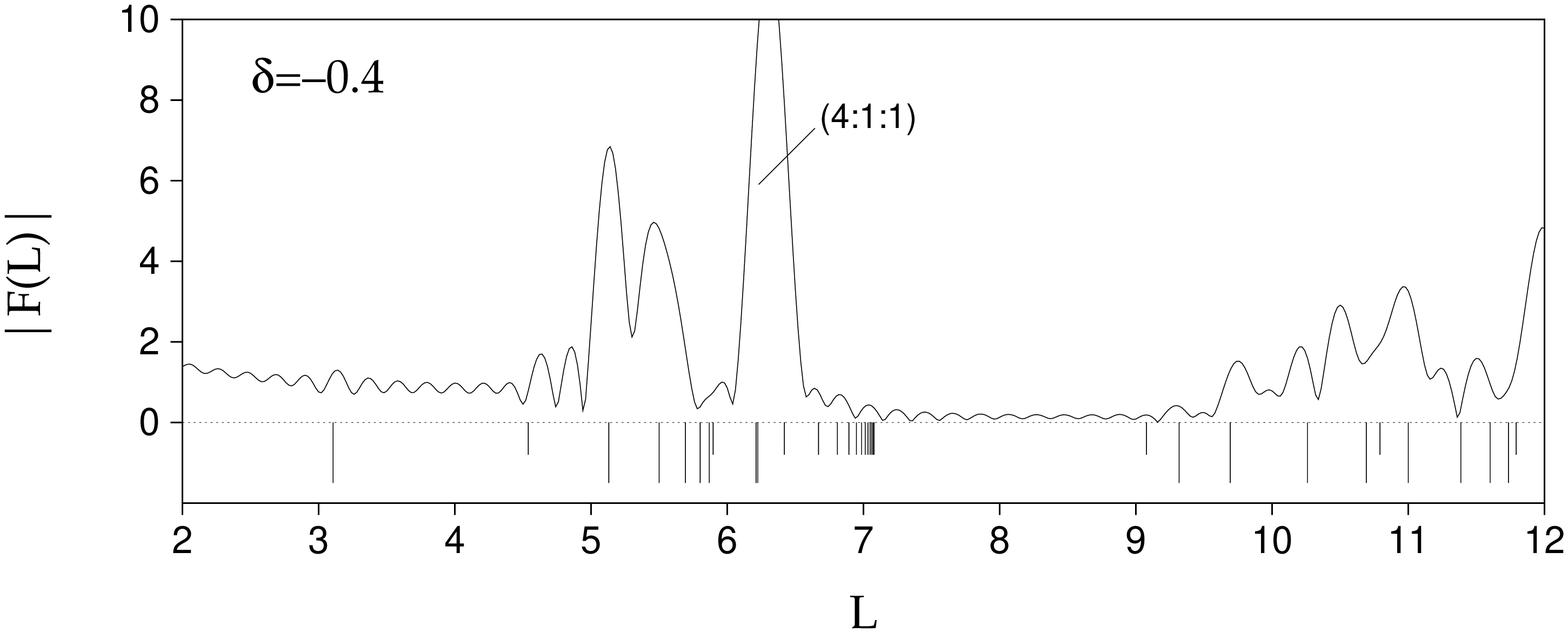}}
\centerline{\epsfxsize=.75\textwidth\epsffile{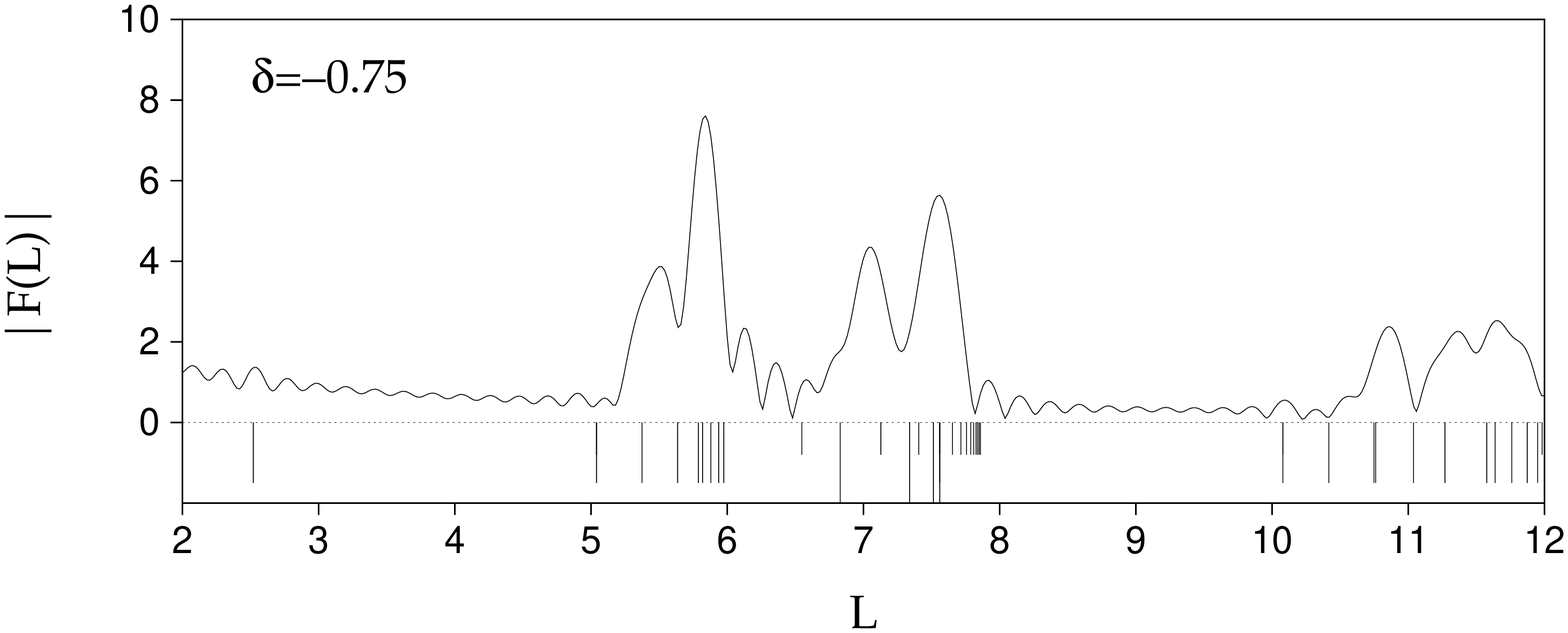}}
\centerline{\epsfxsize=.75\textwidth\epsffile{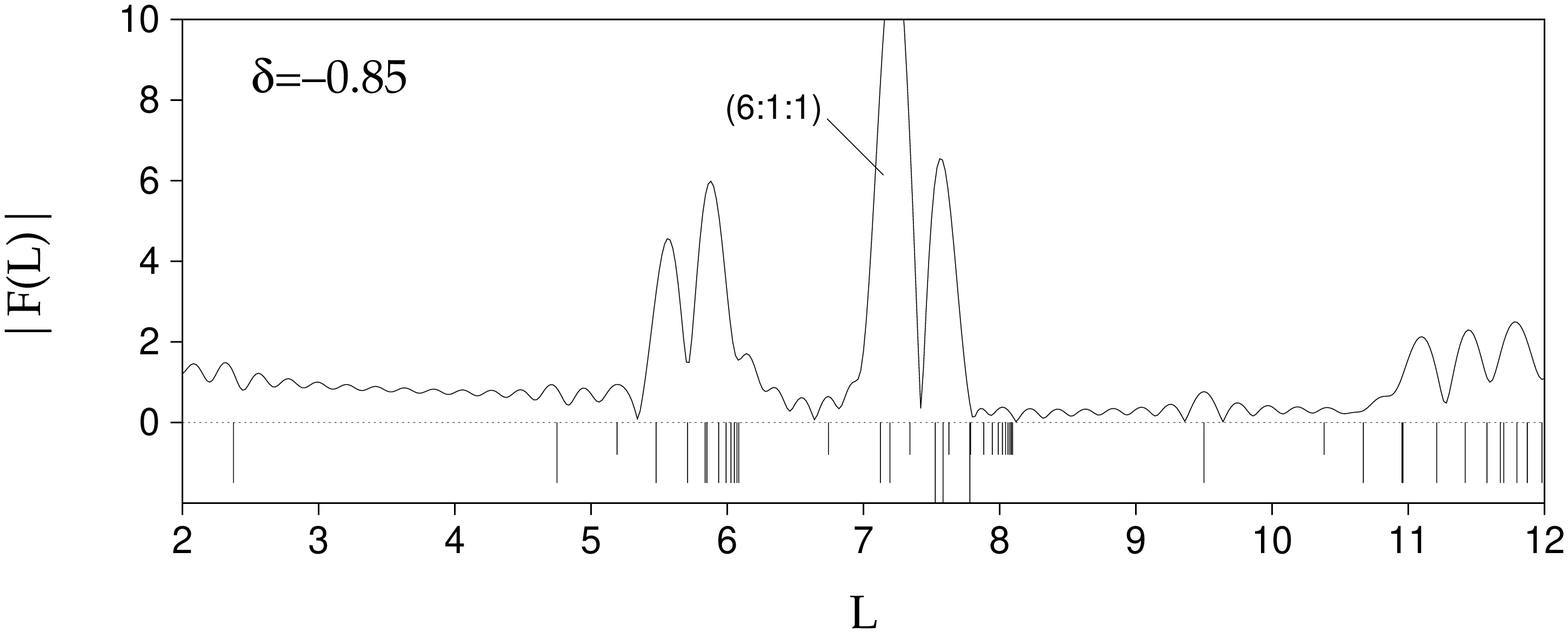}}
\caption{\label{fig:ftl-ob}
Same as Fig.~\protect\ref{fig:ftl-sd} but for oblate cavities with
$\delta=-0.3 \sim -0.85$.}
\end{figure}

Finally let us consider oblate deformations.
Figure~\ref{fig:ftl-ob} displays Fourier transforms of quantum spectra
for oblate spheroidal cavities with $\delta=-0.3\sim -0.85$.
For $\delta=-0.3$, dominant two peaks are associated with triangular
and tetragonal orbits in the meridian plane.  For $\delta=-0.4$, we see a
dominant peak at $L\simeq 6.3$ in addition to the peaks associated
with the meridian-plane orbits.  This new peak is associated with
the butterfly-shaped planar orbits (4:1:1) bifurcated from double
repetitions of linear orbits along the minor axis.
% For $\delta=-0.5$, we see another peak growing up at $L\simeq 6.5$.
% This peak is associated with 3D orbits (4:1:3/2) and (4:1:2), which
% are born simultaneously with the butterfly-shaped planar orbits.

At $\delta=-0.75$ (axis ratio $\eta=2$), the other peak at $L\simeq
7.5$ becomes important.  This peak is associated with the triple
traversals of linear orbits along the minor axis, which bifurcate just
at this shape to the planer hyperbolic orbits (6:1:1).  They make
predominant contribution for $\delta=-0.85$ (the peak at $L\simeq
7.1$).

Constant-action lines for these bifurcated orbits (4:1:1) and (6:1:1)
are indicated in Figs.~\ref{fig:cacmap} and \ref{fig:cacmap2}.  We see
clear correspondence between shapes of these lines and of valleys in
the oscillating level density.  Combining this good correspondence
with the behavior of the Fourier peaks mentioned above, it is evident
that these periodic orbits are responsible for the shell structure at
oblate superdeformations with axis ratio about 2:1.  According to the
classification in section~\ref{sec:bifurcation}, these are W-mode
orbits.

In contrast to W-mode orbits, B-mode 3D orbits do not seem very
important, although those with $\mbox{($p$:$t$:$q$)}=\mbox{(5:1:2)},
\mbox{(6:1:2)},\ldots$ etc. are already bifurcated from equatorial-plane
orbits at the superdeformed region.  This is an important difference
between the prolate and the oblate superdeformations in the spheroidal
cavity model.

\section{Conclusions}
\label{sec:conclusion}

Classical periodic orbits responsible for emergence of the
superdeformed shell structure for single-particle motions in
spheroidal cavities are identified and their relative contributions to
the shell structures are evaluated.  Both prolate and oblate
superdeformations as well as prolate hyperdeformations are
investigated.

Fourier transforms of quantum spectra clearly show that 3D periodic
orbits born out of bifurcations of planar orbits in the equatorial
plane become predominant at large prolate deformations, while
butterfly-shaped planar orbits bifurcated from linear orbits along the
minor axis are important at large oblate deformations.

Good correspondence between constant-action lines for these periodic
orbits and valley structures in the oscillating part of the smoothed
level density confirms the above conclusions.

After writing this paper, we learned that Magner et al.\cite{magner}
had carried out an extensive semiclassical analysis of shell structure
in large prolate cavities.  In their work, a rather large
coarse-graining parameter $\gamma$ for the level density was used, so
that the equatorial-orbit bifurcations discussed in this paper were
not clearly seen.  It remains as a challenge for future to develop a
semiclassical theory capable of treating the equatorial-orbit
bifurcations, and the phase-space trace formula proposed in
Ref.~\citen{magner} seems to provide a general framework for this aim.

\section*{Acknowledgments}

We thank Matthias Brack, Alexander Magner, Zhang Xizhen, Rashid
Nazmitdinov and Masayuki Matsuo for stimulating conversations.


\begin{thebibliography}{99}
\bibitem{twin}
P. J. Nolan and P. J. Twin,
Ann. Rev. Nucl. Part. Sci. {\bf 38} (1988), 533.
\bibitem{khoo}
R. V. F. Janssens and T. L. Khoo,
Ann. Rev. Nucl. Part. Sci. {\bf 41} (1991), 321.
\bibitem{gamma}
{\it Proceedings of the Workshop on Gammasphere Physics},
ed. M. A. Deleplanque, I. Y. Lee and A. O. Macchiavelli
(World Scientific, 1996).
\bibitem{aberg}
S. {\AA}berg, H. Flocard and W. Nazarewicz,
Ann. Rev. Nucl. Part. Sci. {\bf 40} (1990), 439.
\bibitem{gutw}
M. C. Gutzwiller,
J. Math. Phys. {\bf 12} (1971), 343.
\bibitem{balian}
R. Balian and C. Bloch,
Ann. Phys. {\bf 69} (1972), 76.
\bibitem{BT}
M. V. Berry and M. Tabor,
Proc. R. Soc. London {\bf A349} (1976), 101.
\bibitem{BB}
M. Brack and R. K. Bhaduri,
{\it Semiclassical Physics} (Addison-Wesley, Reading, 1997).
\bibitem{stru}
V. M. Strutinsky, A. G. Magner, S. R. Ofengenden and T. D{\o}ssing,
Z. Phys. {\bf A283} (1977), 269.
\bibitem{arvieu1}
R. Arvieu, F. Brut, J. Carbonell and J. Touchard,
Phys. Rev. {\bf A35} (1987), 2389.
\bibitem{arvieu2}
Y. Ayant and R. Arvieu,
J. of Phys. {\bf A20} (1987), 397.
\bibitem{arvieu3}
R. Arvieu and Y. Ayant,
J. of Phys. {\bf A20} (1987), 1115.
\bibitem{frisk}
H. Frisk,
Nucl. Phys. {\bf A511} (1990), 309.
\bibitem{nishi1}
H. Nishioka, M. Ohta and S. Okai,
Mem. Konan Univ. Sci. Ser., {\bf 38}(2) (1991), 1.
\bibitem{nishi2}
H. Nishioka, N. Nitanda, M. Ohta and S. Okai,
Mem. Konan Univ. Sci. Ser., {\bf 39}(2) (1992), 67.
\bibitem{conf}
K.~Arita, A.~Sugita and K.~Matsuyanagi, 
{\it Proc. Int. Symp. on Similarities and Differences between Nuclei and
Clusters}, Tsukuba, July 1--4, 1997,
AIP Conference Proceedings 416, ed. Y.~Abe, I.~Arai, S.~M.~Lee
and K.~Yabana, p.~393. \\
{\it Proc. Int. Symp. on Atomic Nuclei and Metallic Clusters:
Finite Many-Fermion Systems},
Prague, Sep. 1--5, 1997, Czech. J. of Phys. {\bf 48} (1998), 821. \\
{\it Proc. Int. Conf. on Nuclear Structure and Related Topics},
Dubna, Sep. 9--13, 1997,
ed. S.~N.~Ershov, R.~V.~Jolos and V.~V.~Voronov, p.~198.
\bibitem{pal}
T. Mukhopadhyay and S. Pal,
Nucl. Phys. {\bf A592} (1995), 291.
\bibitem{SAM}
A. Sugita, K. Arita and K. Matsuyanagi,
preprint KUNS1431 (1997),
Prog. Theor. Phys. {\bf 100} (1998), in press.
\bibitem{NHM}
H. Nishioka, Klavs Hansen and B. R. Mottelson,
Phys. Rev. {\bf B42} (1990), 9377.
\bibitem{magner}
A.~G.~Magner, S.~N.~Fedotkin, F.~A.~Ivanyuk, P.~Meier, M.~Brack,
S.~M.~Reimann, and H.~Koizumi,
Ann. Physik {\bf 6}, (1997), 555.\\
see also A. G. Magner, S. N. Fedotkin, F. A. Ivanyuk, P. Meier
and M. Brack,
{\it Proc. Int. Conf. on Atomic Nuclei and Metallic Clusters:
Finite Many-Fermion Systems}, Prague, Sep. 1--5, 1997,
Czech. J. of Phys. {\bf 48} (1998), 845.
\end{thebibliography}
\end{document}